\documentstyle[aps,prbbib]{revtex}
\newcommand{\etal}{{\it et al.\/}}
\newcommand{\Exp}[1]{ {e^{#1}} }
\newcommand{\etc}{{\it etc\/}}
\begin{document}
\draft

\title{Calculating photonic Green's functions using
a non-orthogonal finite difference time domain method}
\author{A. J. Ward, J. B. Pendry}
\address{Condensed Matter Theory Group, The Blackett Laboratory,
Imperial College, London, SW7 2BZ, UK}
\date{\today}
\maketitle

\begin{abstract}
In this paper we shall propose a simple scheme for calculating Green's
functions for photons propagating in complex structured dielectrics or
other photonic systems. The method is based on an extension of the
finite difference time domain 
(FDTD) method, originally proposed by Yee~\cite{Yee}, also known as the
Order-N method~\cite{ordern}, which has
recently become a popular way of calculating photonic band structures.
We give a new, transparent derivation of the Order-N method which, in
turn,
enables us to give a simple yet rigorous derivation of the criterion
for numerical stability as well as statements of charge and energy
conservation which are exact even on the discrete lattice.
We implement this using a general, non-orthogonal co-ordinate system
without incurring the computational overheads normally associated with
non-orthogonal FDTD.

We present results for local densities of states calculated using this
method for a number of systems.
Firstly, we consider a simple one dimensional dielectric multilayer,
identifying the suppression in the state density caused by the photonic
band gap and then observing the effect of introducing a defect layer
into the periodic structure. 
Secondly, we tackle a more realistic example by treating a
defect in a crystal of dielectric spheres on a diamond lattice.
This could have application to the design of super-efficient laser
devices utilising defects in photonic crystals as laser cavities.
\end{abstract}


\section{Introduction}

One of the principle driving forces behind the recent flurry of
research into photonic band gap materials~\cite{JBPrev,Yabrev1} has
been the potential for 
manipulating the spontaneous emission of an atom placed in a cavity in
such a material. As was pointed out some years ago~\cite{Yab1} a
material with a periodically structured dielectric function (a
photonic crystal) can have a
profound effect on the density of states for photons within the
material, in some cases leading to frequency windows for which no
allowed photon states exist. These windows, or photonic band gaps,
radically alter the emission properties of atoms. An excited atom that
wants to emit a photon of a frequency within the band gap cannot do
so---the photon has no states into which it can go and so is
forced to from a new kind of atom-photon bound state. Even zero-point
fluctuations are forbidden within the band gap.
This has immediate implications for device physics. Placing an active
device, such as a semiconductor laser, within a cavity in a photonic
crystal offers the possibility to control unwanted spontaneous
emission and allow emission only into the lasing mode, thus
dramatically improving the efficiency of the device~\cite{japan}.

Green's functions have the potential to play a central role in the
theoretical investigations of these photonic systems. Not only are
they a natural way to express key quantities such as the density of
states, they are also are easily calculated within the framework of
time domain methods such as the Order~N technique that we shall be
exploring in this paper. In systems where dissipation is present, such
as those containing metals or lossy dielectrics, the Green's function
approach is the only route to calculate quantities of physical interest.

For the theorist, the challenge is to solve Maxwell's equations for
such systems. Much progress has been made in the past few years and
several well established techniques have emerged. Probably the most
widely used is the Plane Wave method~\cite{Leu,Zha,Ho}. Simply put,
this method involves expanding the electromagnetic fields as a sum of
plane waves and recasting Maxwell's equations into the form of a
eigenvalue problem to find the allowed eigenfrequencies. Though simple
to implement and use, this method 
has the draw back that the time taken for the calculation scales as
the cube of the number of plane waves used, so for complicated problems
which require many plane waves this is a severe limitation. 
Another limitation is to systems whose
dielectric functions do not disperse with frequency. Hence metallic
systems are beyond the scope of this method.

A second popular method involves working at a fixed frequency but
instead of expanding the wavefield on a lattice in reciprocal space
the wavefield is represented on the points of a real space
lattice~\cite{J+A,JBP}. 
The resulting equations can be rearranged into the form of a
Transfer Matrix which relates the fields in one
layer of the lattice to the fields in the next. This method has proved
extremely useful especially for systems involving metals where the
dielectric constant is a function of frequency. Also, 
because the form the transfer
matrix takes, connecting the fields on one {\em surface} to the
fields on another, calculations based on this method scale as the
square of the number of real space points, rather than the cube. 

This may be an improvement over
the plane wave method but is still worse than the optimal linear
scaling with system size. However, it has been shown~\cite{ordern}
that, because Maxwell's equations are local then by working in the
time domain instead of the 
frequency domain it is possible to obtain methods which scale as
`order $N$' where $N$ is the system size. We shall work with such a
method in this paper in order to calculate photonic Green's
functions and from those other quantities of physical interest such as
densities of states. Previous work has successfully calculated spontaneous
emission rates from the density of states using a real space lattice
formulation~\cite{frank}. However, that work was based in the frequency
domain. By working in the time domain we are able to exploit the
more favourable scaling law and consider larger, more complex systems
such as photonic crystals with defects.

\section{Methods}

The theory behind the methods used in this paper is somewhat
similar in principle to the Finite Difference Time Domain method
first introduced to the electrical engineering community by Yee~\cite{Yee}
in 1966, and first applied to the problem of photonic band structures by Chan
\etal~\cite{ordern} in 1995. 
Here we present a systematic derivation of the finite difference
equations within a transparent formalism.
The advantages of this new
formalism over the traditional Yee approach are many. First, it makes
clear how to find quantities which are exactly conserved even by the
discrete equations. Hence we can identify the analogies to charge,
energy density and the Poynting vector \etc. Second, our formalism
enables the precise analysis of the stability of the discrete
equations to be
made from a simple consideration of the approximations involved,
sidestepping the usual, rather involved, Courant stability analysis.
Third, and perhaps most importantly, the new formalism shows, for
the first time, how to present the finite difference equations in a
completely general co-ordinate system without incurring the
computational overheads normally associated with non-orthogonal
FDTD~\cite{mittra}.
Finally, it is hoped that this transparency will make it simpler to extent
the method to areas to which it has not yet been applied.
We begin from the usual Maxwell's equations,
neglecting any free charges or currents;
\begin{equation}
\nabla\times{\bf H}=\varepsilon_{0}\varepsilon({\bf r})
\frac{\partial{\bf E}}{\partial t}
\hspace{10mm}
\nabla\times{\bf E}=-\mu_{0}\mu({\bf r})\frac{\partial{\bf H}}
{\partial t} 
\end{equation}
which on Fourier transforming into $({\bf K},\omega)$ space can be
written,
\begin{equation}
i{\bf k}\times{\bf H}=-i\varepsilon_{0}\varepsilon
\omega{\bf E}
\hspace{1cm}
i{\bf k}\times{\bf E}=+i\mu_{0}\mu
\omega{\bf H}
\label{eq:max}
\end{equation}

Next we wish to place this equations onto a discrete, real space
lattice by replacing the derivatives with
finite differences. We have some freedom here but must be
careful---not all differencing schemes lead to stable equations. We
will take our lead therefore from the differencing scheme
which has proved so successful in the transfer matrix
method~\cite{JBP}. 
We do this by introducing 
the following approximations to $\omega$ and ${\bf k}$. For the
terms which involve the electric field we use,
\begin{equation}
k_x\mapsto k^{+}_x=\frac{\Exp{(ik_x a)}-1}{ia}
\hspace{1cm}
\omega\mapsto \omega^{+}=\frac{\Exp{(-i\omega\delta t)}-1}{-i\delta t}
\label{eqn:approx1}
\end{equation}
with similar expressions for $k^{+}_y$ and $k^{+}_z$. For the magnetic field
terms we use,
\begin{equation}
k_x\mapsto k^{-}_x=\frac{1-\Exp{(-ik_x a)}}{ia}
\hspace{1cm}
\omega\mapsto \omega^{-}=\frac{1-\Exp{(i\omega\delta t)}}{-i\delta t}
\label{eqn:approx2}
\end{equation}
\etc. So making these approximations  in equation~(\ref{eq:max}) we get,
\begin{equation}
i{\bf k}^{-}\times{\bf H}=-i\varepsilon_{0}\varepsilon
\omega^{+}{\bf E}
\hspace{1cm}
i{\bf k}^{+}\times{\bf E}=+i\mu_{0}\mu
\omega^{-}{\bf H}
\end{equation}
%Or writing it out in full,
%\begin{equation}
%-\varepsilon_0\varepsilon({\bf r})
%\frac{\Exp{[-i\omega\delta t]}-1}{-i\delta t}E_x({\bf r},t)=
%\frac{1-\Exp{[-ik_z c]}}{ic}H_{z}({\bf r},t) - 
%\frac{1-\Exp{[-ik_y b]}}{ib}H_{y}({\bf r},t) 
%\end{equation}
%\begin{equation}
%-\varepsilon_0\varepsilon({\bf r})
%\frac{\Exp{[-i\omega\delta t]}-1}{-i\delta t}E_y({\bf r},t)=
%\frac{1-\Exp{[-ik_x a]}}{ia}H_{x}({\bf r},t) - 
%\frac{1-\Exp{[-ik_z c]}}{ic}H_{z}({\bf r},t) 
%\end{equation}
%\begin{equation}
%-\varepsilon_0\varepsilon({\bf r})
%\frac{\Exp{[-i\omega\delta t]}-1}{-i\delta t}E_z({\bf r},t)=
%\frac{1-\Exp{[-ik_y b]}}{ib}H_{y}({\bf r},t) - 
%\frac{1-\Exp{[-ik_x a]}}{ia}H_{x}({\bf r},t) 
%\end{equation}
%\begin{equation}
%\mu_0\mu({\bf r})
%\frac{1-\Exp{[i\omega\delta t]}}{-i\delta t}H_x({\bf r},t)=
%\frac{\Exp{[ik_z c]}-1}{ic}E_{z}({\bf r},t) - 
%\frac{\Exp{[ik_y b]}-1}{ib}E_{y}({\bf r},t) 
%\end{equation}
%\begin{equation}
%\mu_0\mu({\bf r})
%\frac{1-\Exp{[i\omega\delta t]}}{-i\delta t}H_y({\bf r},t)=
%\frac{\Exp{[ik_x a]}-1}{ia}E_{x}({\bf r},t) - 
%\frac{\Exp{[ik_z c]}-1}{ic}E_{z}({\bf r},t) 
%\end{equation}
%\begin{equation}
%\mu_0\mu({\bf r})
%\frac{1-\Exp{[i\omega\delta t]}}{-i\delta t}H_z({\bf r},t)=
%\frac{\Exp{[ik_y b]}-1}{ib}E_{z}({\bf r},t) - 
%\frac{\Exp{[ik_x a]}-1}{ia}E_{y}({\bf r},t) 
%\end{equation}
On Fourier transforming back into the $({\bf r},t)$ domain it is clear
that these approximations are equivalent to taking a forward finite
difference, $\Delta^{+}$, in place of derivatives of the electric
field, and a backwards 
difference, $\Delta^{-}$, for derivatives of the magnetic field.
\begin{equation}
\Delta^{+}_{x}{\cal F}({\bf r})=
\left [{\cal F}({\bf r+a}) -{\cal F}({\bf r})
\right ] / a
\end{equation}
\begin{equation}
\Delta^{-}_{x}{\cal F}({\bf r})=
\left [{\cal F}({\bf r}) -{\cal F}({\bf r-a})
\right ] / a 
\end{equation}
%Similarly we
%replace derivatives in the $y$ and $z$-directions with operators
%$\Delta^{\pm}_{y}$ and $\Delta^{\pm}_{z}$. 
Putting it all together the discrete form of Maxwell's equations
become, 
\begin{equation}
\nabla^{+}\times{\bf E}({\bf r},t)=-\mu_{0}\mu({\bf r})
\Delta^{-}_{t} {\bf H}({\bf r},t)
\end{equation}
\begin{equation}
\nabla^{-}\times{\bf H}({\bf r},t)=\varepsilon_{0}
\varepsilon({\bf r})\Delta^{+}_{t} {\bf E}({\bf r},t)
\end{equation}
where,
\begin{equation}
\nabla^{+}\times=
\left ( \begin{array}{ccc}
0 & -\Delta^{+}_{z} & \Delta^{+}_{y} \\
\Delta^{+}_{z} & 0 & -\Delta^{+}_{x} \\
-\Delta^{+}_{y} & \Delta^{+}_{x} & 0 \\
\end{array} \right )
\end{equation}
%\hspace{1mm};\hspace{1mm}
\begin{equation}
\nabla^{-}\times=
\left ( \begin{array}{ccc}
0 & -\Delta^{-}_{z} & \Delta^{-}_{y} \\
\Delta^{-}_{z} & 0 & -\Delta^{-}_{x} \\
-\Delta^{-}_{y} & \Delta^{-}_{x} & 0 \\
\end{array} \right )
\end{equation}
%At this stage it is possible to introduce a transformation to a
%generalised co-ordinate system $(x,y,z)\rightarrow(q_{1},q_{2},q_{3})$
The approximations outlined in the previous paragraph place Maxwell's
equations onto a discrete lattice of points which is uniform and
Cartesian. However, this is not always convenient for the problems we may
want to consider. It may be to our advantage to work in a co-ordinate system
which is non-uniform or even non-orthogonal. Fortunately, as shown in detail
elsewhere~\cite{AJW2} there is a simple result which allows us to map a
completely arbitrary co-ordinate system onto a uniform Cartesian one as long
as we introduce new, renormalised versions of the permittivity and
permeability. In the generalised co-ordinates Maxwell's equations become,
\begin{equation}
\frac{\nabla^{-}_{q}\times \hat{{\bf H}}({\bf r},t)}{Q_0}
=\varepsilon_0\hat{\varepsilon}({\bf r})\frac{\Delta^{+}_{\tau}
\hat{{\bf E}}({\bf r},t)}{\delta t}
\end{equation}
\begin{equation}
\frac{\nabla^{+}_{q}\times \hat{{\bf E}}({\bf r},t)}{Q_0}=
-\mu_0\hat{\mu}({\bf r})
\frac{\Delta^{-}_{\tau}\hat{{\bf H}}({\bf r},t)}{\delta t}
\end{equation}
where,
\begin{equation}
\Delta^{+}_{q_1}{\cal F}({\bf r},t)=
{\cal F}({\bf r}+Q_{1}{\bf u_{1}},t) -{\cal F}({\bf r},t)
\hspace{1cm}
\Delta^{+}_{\tau}{\cal F}({\bf r},t)=
{\cal F}({\bf r},t+\delta t) -{\cal F}({\bf r},t)
\end{equation}
\begin{equation}
\Delta^{-}_{q_1}{\cal F}({\bf r},t)=
{\cal F}({\bf r},t) -{\cal F}({\bf r}-Q_{1}{\bf u_{1}},t)
\hspace{1cm}
\Delta^{-}_{\tau}{\cal F}({\bf r},t)=
{\cal F}({\bf r},t) -{\cal F}({\bf r},t-\delta t)
\end{equation}
etc,
\begin{equation}
\hat{\varepsilon}^{ij}({\bf r})=\varepsilon({\bf r})
g^{ij}|{\bf u_{1}\cdot u_{2}\times u_{3}}|\frac{Q_{1}Q_{2}Q_{3}}
{Q_{i}Q_{j}Q_{0}}
\hspace{1cm}
\hat{\mu}^{ij}({\bf r})=\mu({\bf r})
g^{ij}|{\bf u_{1}\cdot u_{2}\times u_{3}}|\frac{Q_{1}Q_{2}Q_{3}}
{Q_{i}Q_{j}Q_{0}}
\label{eqn:renorm1}
\end{equation}
and
\begin{equation}
\hat{\sigma}^{ij}({\bf r})=\sigma({\bf r})
g^{ij}|{\bf u_{1}\cdot u_{2}\times u_{3}}|\frac{Q_{1}Q_{2}Q_{3}}
{Q_{i}Q_{j}Q_{0}}
\hspace{1cm}
\hat{\sigma}_{m}^{ij}({\bf r})=\sigma_m({\bf r})
g^{ij}|{\bf u_{1}\cdot u_{2}\times u_{3}}|\frac{Q_{1}Q_{2}Q_{3}}
{Q_{i}Q_{j}Q_{0}}
\end{equation}
The generalised co-ordinate system is defined by the three unit vectors,
${\bf u}_1$, ${\bf u}_2$ and ${\bf u}_3$ which point along the
generalised co-ordinate axis and the three lattice spacings $Q_1$, $Q_2$ and
$Q_3$ which define the spacing between discrete lattice points in each
direction. In general, the ${\bf u}_i$'s and $Q_i$'s will themselves by
functions of position. The metric tensor $g^{ij}$ is defined such that
$(g^{-1})^{ij}={\bf u}_{i}\cdot{\bf u}_{j}$.
As well as the permittivity and permeability being renormalised by the
co-ordinate transformation, the fields themselves are also rescaled by the
lattice spacings,
\begin{equation}
\hat{E}_{i}=Q_{i}E_{i}\hspace{2cm}
\hat{H}_{i}=Q_{i}H_{i}
\end{equation}

In order to obtain the equations that will link the fields at one time step
to the fields at the next we introduce,
\begin{equation}
\hat{{\bf H}}'=\frac{\delta t}{\varepsilon_{0}Q_{0}}\hat{{\bf H}}
\end{equation}
and let,
\begin{equation}
{\bf a}=Q_1{\bf u}_1 \hspace{1cm}
{\bf b}=Q_2{\bf u}_2 \hspace{1cm}
{\bf c}=Q_3{\bf u}_3
\end{equation} 
Then after some rearranging we arrive at,
\begin{eqnarray}
\hat{E}_{1}({\bf r},t+\delta t)=
\hat{E}_{1}({\bf r},t)
&+&\left[\hat{\varepsilon}^{-1}
({\bf r})\right]^{11}\left[
 \hat{H}'_{3}({\bf r},t)-\hat{H}'_{3}({\bf r-b},t)
-\hat{H}'_{2}({\bf r},t)+\hat{H}'_{2}({\bf r-c},t)\right]\nonumber\\
&+&\left[\hat{\varepsilon}^{-1}
({\bf r})\right]^{12}\left[
 \hat{H}'_{1}({\bf r},t)-\hat{H}'_{1}({\bf r-c},t)
-\hat{H}'_{3}({\bf r},t)+\hat{H}'_{3}({\bf r-a},t)\right]\nonumber\\
&+&\left[\hat{\varepsilon}^{-1}
({\bf r})\right]^{13}\left[
 \hat{H}'_{2}({\bf r},t)-\hat{H}'_{2}({\bf r-a},t)
-\hat{H}'_{1}({\bf r},t)+\hat{H}'_{1}({\bf r-b},t)\right]\nonumber\\
\label{eq:on1}
\end{eqnarray}
\begin{eqnarray}
\hat{E}_{2}({\bf r},t+\delta t)=
\hat{E}_{2}({\bf r},t)
&+&\left[\hat{\varepsilon}^{-1}
({\bf r})\right]^{21}\left[
 \hat{H}'_{3}({\bf r},t)-\hat{H}'_{3}({\bf r-b},t)
-\hat{H}'_{2}({\bf r},t)+\hat{H}'_{2}({\bf r-c},t)\right]\nonumber\\
&+&\left[\hat{\varepsilon}^{-1}
({\bf r})\right]^{22}\left[
 \hat{H}'_{1}({\bf r},t)-\hat{H}'_{1}({\bf r-c},t)
-\hat{H}'_{3}({\bf r},t)+\hat{H}'_{3}({\bf r-a},t)\right]\nonumber\\
&+&\left[\hat{\varepsilon}^{-1}
({\bf r})\right]^{23}\left[
 \hat{H}'_{2}({\bf r},t)-\hat{H}'_{2}({\bf r-a},t)
-\hat{H}'_{1}({\bf r},t)+\hat{H}'_{1}({\bf r-b},t)\right]\nonumber\\
\end{eqnarray}
\begin{eqnarray}
\hat{E}_{3}({\bf r},t+\delta t)=
\hat{E}_{3}({\bf r},t)
&+&\left[\hat{\varepsilon}^{-1}
({\bf r})\right]^{31}\left[
 \hat{H}'_{3}({\bf r},t)-\hat{H}'_{3}({\bf r-b},t)
-\hat{H}'_{2}({\bf r},t)+\hat{H}'_{2}({\bf r-c},t)\right]\nonumber\\
&+&\left[\hat{\varepsilon}^{-1}
({\bf r})\right]^{32}\left[
 \hat{H}'_{1}({\bf r},t)-\hat{H}'_{1}({\bf r-c},t)
-\hat{H}'_{3}({\bf r},t)+\hat{H}'_{3}({\bf r-a},t)\right]\nonumber\\
&+&\left[\hat{\varepsilon}^{-1}
({\bf r})\right]^{33}\left[
 \hat{H}'_{2}({\bf r},t)-\hat{H}'_{2}({\bf r-a},t)
-\hat{H}'_{1}({\bf r},t)+\hat{H}'_{1}({\bf r-b},t)\right]\nonumber\\
\end{eqnarray}
\begin{eqnarray}
\hat{H}'_{1}({\bf r},t+\delta t)&=&
\hat{H}'_{1}({\bf r},t)\nonumber\\
&-&\left(\frac{\delta t\ c_{0}}{Q_{0}}\right)^{2}
\left[\hat{\mu}^{-1}({\bf r})\right]^{11}\left[
 \hat{E}_{3}({\bf r+b},t)-\hat{E}_{3}({\bf r},t)
-\hat{E}_{2}({\bf r+c},t)+\hat{E}_{2}({\bf r},t)\right]\nonumber\\
&-&\left(\frac{\delta t\ c_{0}}{Q_{0}}\right)^{2}
\left[\hat{\mu}^{-1}({\bf r})\right]^{12}\left[
 \hat{E}_{1}({\bf r+c},t)-\hat{E}_{1}({\bf r},t)
-\hat{E}_{3}({\bf r+a},t)+\hat{E}_{3}({\bf r},t)\right]\nonumber\\
&-&\left(
\frac{\delta t\ c_{0}}{Q_{0}}\right)^{2}
\left[\hat{\mu}^{-1}({\bf r})\right]^{13}\left[
 \hat{E}_{2}({\bf r+a},t)-\hat{E}_{2}({\bf r},t)
-\hat{E}_{1}({\bf r+b},t)+\hat{E}_{1}({\bf r},t)\right]
\end{eqnarray}
\begin{eqnarray}
\hat{H}'_{2}({\bf r},t+\delta t)&=&
\hat{H}'_{2}({\bf r},t)\nonumber\\
&-&\left(\frac{\delta t\ c_{0}}{Q_{0}}\right)^{2}
\left[\hat{\mu}^{-1}({\bf r})\right]^{21}\left[
 \hat{E}_{3}({\bf r+b},t)-\hat{E}_{3}({\bf r},t)
-\hat{E}_{2}({\bf r+c},t)+\hat{E}_{2}({\bf r},t)\right]\nonumber\\
&-&\left(\frac{\delta t\ c_{0}}{Q_{0}}\right)^{2}
\left[\hat{\mu}^{-1}({\bf r})\right]^{22}\left[
 \hat{E}_{1}({\bf r+c},t)-\hat{E}_{1}({\bf r},t)
-\hat{E}_{3}({\bf r+a},t)+\hat{E}_{3}({\bf r},t)\right]\nonumber\\
&-&\left(
\frac{\delta t\ c_{0}}{Q_{0}}\right)^{2}
\left[\hat{\mu}^{-1}({\bf r})\right]^{23}\left[
 \hat{E}_{2}({\bf r+a},t)-\hat{E}_{2}({\bf r},t)
-\hat{E}_{1}({\bf r+b},t)+\hat{E}_{1}({\bf r},t)\right]
\end{eqnarray}
\begin{eqnarray}
\hat{H}'_{3}({\bf r},t+\delta t)&=&
\hat{H}'_{3}({\bf r},t)\nonumber\\
&-&\left(\frac{\delta t\ c_{0}}{Q_{0}}\right)^{2}
\left[\hat{\mu}^{-1}({\bf r})\right]^{31}\left[
 \hat{E}_{3}({\bf r+b},t)-\hat{E}_{3}({\bf r},t)
-\hat{E}_{2}({\bf r+c},t)+\hat{E}_{2}({\bf r},t)\right]\nonumber\\
&-&\left(\frac{\delta t\ c_{0}}{Q_{0}}\right)^{2}
\left[\hat{\mu}^{-1}({\bf r})\right]^{32}\left[
 \hat{E}_{1}({\bf r+c},t)-\hat{E}_{1}({\bf r},t)
-\hat{E}_{3}({\bf r+a},t)+\hat{E}_{3}({\bf r},t)\right]\nonumber\\
&-&\left(
\frac{\delta t\ c_{0}}{Q_{0}}\right)^{2}
\left[\hat{\mu}^{-1}({\bf r})\right]^{33}\left[
 \hat{E}_{2}({\bf r+a},t)-\hat{E}_{2}({\bf r},t)
-\hat{E}_{1}({\bf r+b},t)+\hat{E}_{1}({\bf r},t)\right]
\label{eq:on2}
\end{eqnarray}
These equations allow us to take an arbitrary set of electric and
magnetic fields at some initial time $t=0$ and, subject to appropriate
boundary conditions, calculate the fields at all later times.
%This in turn, permits us to calculate all the quantities we may
%want. For example, if our system is periodic and subject to Bloch
%boundary conditions corresponding to a particular ${\bf k}$, then the
%Fourier transform 
%of the time dependent fields into the frequency domain allows us to
%identify the eigenfrequencies for that ${\bf k}$, and hence build
%up the photonic band structure. 

Notice that by incorporating all of the details of the generalised
co-ordinate system into the new definitions of the permittivity and
permeability we have, in effect, eliminated the extra computational
overhead of the general co-ordinates. If the co-ordinate system
changed from one place to another, and there is no reason why it
should not, we would have to store the metric tensor at each point on
the lattice. But since we have to store $\varepsilon$ and $\mu$
tensors at
each point anyway, by combining the metric into our new
$\hat{\varepsilon}$ and $\hat{\mu}$ we cut the amount of storage
needed. Similarly, we also reduce the number of calculations required
at each time step as we no longer need to worry about converting
between covariant and contravariant vectors at every time step - this
is all taken care 
of by $\hat{\varepsilon}$ and $\hat{\mu}$. Hence, in contrast with
previous methods, our non-orthogonal FDTD has no additional
computational overhead compared to an orthogonal one, except for the
initial setting up of $\hat{\varepsilon}$ and $\hat{\mu}$.

\subsection{Stability Criterion}
These equations give a stable updating procedure for the fields if
the time step is kept sufficiently small. The criterion is easy to find.
Starting from the approximations we made for ${\bf k}$ and
$\omega$, (equations~\ref{eqn:approx1} and~\ref{eqn:approx2}) the free
space dispersion relation, 
$\omega^{2}=c_{0}^{2}k^{2}$, becomes,
\begin{equation}
\frac{4}{\delta t^2}\sin^2\left(\frac{\omega \delta t}{2}\right)=
4c_0^2\left[
 \frac{1}{Q_1^2}\sin^2\left(\frac{Q_1\ k_x}{2}\right)
+\frac{1}{Q_2^2}\sin^2\left(\frac{Q_2\ k_y}{2}\right)
+\frac{1}{Q_3^2}\sin^2\left(\frac{Q_3\ k_z}{2}\right)\right]
\end{equation}
The condition that the maximum value of the right hand side must
correspond to a real frequency gives,
\begin{equation}
(\delta t)^2<\left(\frac{c_0^2}{Q_1^2}+\frac{c_0^2}{Q_2^2}
+\frac{c_0^2}{Q_3^2}\right)^{-1}
\end{equation}

\subsection{Fourier Transforms}

The final step of any time domain method is a Fourier transform of the
time dependent information into the frequency domain, and for our FDTD
method there are a few points which need to be made clear. Firstly, it
is desirable to eliminate the zero frequency, longitudinal mode from
our results. This is done by subtracting off the static part from the
time dependent fields which we have calculated so that their time average
is zero. Next, we must ensure that we obtain the properly causal
solution to Maxwell's equations when we replace the integral in the
Fourier transform with a discrete sum. We do this by
adding a small positive, imaginary part $\delta$ to the frequency.
\begin{equation}
\int_{-\infty}^{\infty}f(t)\exp{[i\omega\,t]}\,dt
\mapsto\sum_{n=1}^{N_{t}} f(n\,\delta t)
\exp{[i(\omega+i\delta)\,(n\pm1/2)\delta t]} 
\end{equation}
The size of
this imaginary part is determined by the total time interval over
which the fields are integrated. In order to be properly causal, the
final term in the sum must tend to zero. This is of less
importance when finding a band structure but is critical
if we are to calculate the Green's function correctly. Another small
detail - because of the choices we have made in our approximations for
the time derivatives for ${\bf E}$ and ${\bf H}$, it is necessary
to include a half time step offset, minus for the ${\bf E}$ field
and plus for the ${\bf H}$.

\subsection{Conserved Quantities}
\label{sec:cons}

An obvious question to ask is whether our discrete
Maxwell's equations have conserved quantities analogous to those for
the continuum equations. The simplest to start with is the
conservation of charge.
If we consider the quantity, 
\begin{equation}
\nabla_{q}^{-}\cdot\hat{\varepsilon}({\bf r})
{\bf \hat{E}(r)}
\end{equation}
And then calculate the discrete version of its time derivative,
\begin{eqnarray}
\Delta_{\tau}^{-}\cdot\nabla_{q}^{-}\cdot\hat{\varepsilon}({\bf r})
{\bf \hat{E}(r)}
&=&\nabla_{q}^{-}\cdot\Delta_{\tau}^{-}\cdot\hat{\varepsilon}({\bf r})
{\bf \hat{E}(r)}\nonumber\\
&=&\nabla_{q}^{-}\cdot\nabla_{q}^{-}\times\hat{{\bf H}}'\nonumber\\
&=&0
\end{eqnarray}
Because our approximations to Maxwell's equations have preserved the
form of the curls, the quantity $\nabla_{q}^{-}\cdot\hat{\varepsilon}
({\bf r}){\bf \hat{E}(r)}$ remains an exactly conserved quantity
even on the discrete lattice. This is the direct equivalent of
$\nabla\cdot{\bf D}$ being conserved in the continuum case, in
other words, of the conservation of charge.
The same follows for  $\nabla^{+}_{q}\cdot\hat{\mu}({\bf r})
\hat{{\bf H}}'({\bf r})$ corresponding to the conservation of the
`magnetic charge', $\nabla\cdot{\bf B}$.

The next task is to find the correct form for Poynting's theorem on
the lattice. This again follows in a straightforward way in our
formalism if we begin from the following form for the energy density.
\begin{equation}
U(t)=\frac{1}{2}\left[
\varepsilon_{0}\varepsilon({\bf r})E_{\alpha}^{*}({\bf r},t)
E^{\alpha}({\bf r},t-\delta t)
+\mu_{0}\mu({\bf r})H_{\alpha}^{*}({\bf r},t-\delta
t)H^{\alpha}({\bf r},t-\delta t) 
\right]
\end{equation}
This has the correct form for the energy density in the limit $\delta t
\rightarrow 0$.
Substituting the expressions for the generalised
co-ordinates we obtain,
\begin{equation}
U(t)=\frac{U_{0}}{2}\left[
\hat{\varepsilon}^{\alpha\beta}\hat{E}^{*}_{\alpha}({\bf r},t)
\hat{E}_{\beta}({\bf r},t-\delta t)
+(\frac{Q_{0}}{c_{0}\delta t})^{2}
\hat{\mu}^{\alpha\beta}\hat{H'}^{*}_{\alpha}({\bf r},t-\delta t)
\hat{H'}_{\beta}({\bf r},t-\delta t)
\right]
\end{equation}
where $ U_{0}=Q_{0}\varepsilon_{0}/(Q_{1}Q_{2}Q_{3}|{\bf u_{1}\cdot
u_{2}\times u_{3}}|)$.
Poynting's theorem can then be obtained by considering
the time difference operator
$\Delta_{t} U(t)=(U(t+\delta t)-U(t))/\delta t$. This gives,
\begin{eqnarray}
\Delta^{+}_{t}U(t)&=&\frac{U_{0}}{2}
\left\{\left[\hat{H}'^{*}_{2}({\bf r-c},t)
-\hat{H}'^{*}_{3}({\bf r-b},t)\right]\hat{E}_{1}({\bf r},t) 
\right.\nonumber\\
&+&\left.\left[\hat{H}'^{*}_{3}({\bf r-a},t)
-\hat{H}'^{*}_{1}({\bf r-c},t)\right]\hat{E}_{2}({\bf r},t) 
\right.\nonumber\\
&+&\left.\left[\hat{H}'^{*}_{1}({\bf r-b},t)
-\hat{H}'^{*}_{2}({\bf r-a},t)\right]\hat{E}_{3}({\bf r},t) 
\right.\nonumber\\
&+&\left.\left[\hat{H}'_{2}({\bf r-c},t-\delta t)
-\hat{H}'_{3}({\bf r-b},t-\delta t)\right]
\hat{E}^{*}_{1}({\bf r},t) \right.\nonumber\\
&+&\left.\left[\hat{H}'_{3}({\bf r-a},t-\delta t)
-\hat{H}'_{1}({\bf r-c},t-\delta t)\right]
\hat{E}^{*}_{2}({\bf r},t) \right.\nonumber\\
&+&\left.\left[\hat{H}'_{1}({\bf r-b},t-\delta t)
-\hat{H}'_{2}({\bf r-a},t-\delta t)\right]
\hat{E}^{*}_{3}({\bf r},t) \right.\nonumber\\
&+&\left.\left[\hat{E}_{2}({\bf r+c},t)-\hat{E}_{3}({\bf r+b},t)\right]
\hat{H}'^{*}_{1}({\bf r},t) \right.\nonumber\\
&+&\left.\left[\hat{E}_{3}({\bf r+a},t)-\hat{E}_{1}({\bf r+c},t)\right]
\hat{H}'^{*}_{2}({\bf r},t) \right.\nonumber\\
&+&\left.\left[\hat{E}_{1}({\bf r+b},t)-\hat{E}_{2}({\bf r+a},t)\right]
\hat{H}'^{*}_{3}({\bf r},t) \right.\nonumber\\
&+&\left.\left[\hat{E}^{*}_{2}({\bf r+c},t)
-\hat{E}^{*}_{3}({\bf r+b},t)\right]
\hat{H}'_{1}({\bf r},t) \right.\nonumber\\
&+&\left.\left[\hat{E}^{*}_{3}({\bf r+a},t)
-\hat{E}^{*}_{1}({\bf r+c},t)\right]
\hat{H}'_{2}({\bf r},t) \right.\nonumber\\
&+&\left.\left[\hat{E}^{*}_{1}({\bf r+b},t)
-\hat{E}^{*}_{2}({\bf r+a},t)\right]
\hat{H}'_{3}({\bf r},t)
\right\}
\label{eq:poynting}
\end{eqnarray}
The important point to notice is that when this quantity is summed
over a set of neighbouring lattice points only the terms associated
with the surface of the integration region survive, all the volume
terms cancel. This allows us to identify the right hand side of
equation~(\ref{eq:poynting}) as the Poynting vector integrated over the
volume surrounding one lattice point.

\section{Obtaining the Green's function on the lattice}

We turn our attention now to consider how we can define the Green's
function within our discrete real space formulism. We will begin from
the continuum limit, and write Maxwell's equations as
\begin{equation}
{\bf M}\left(\begin{array}{c}{\bf E} \\ {\bf H}\end{array}\right)
=\omega{\bf P}\left(\begin{array}{c}{\bf E} \\ {\bf H}\end{array}
\right)
\label{eqn:dos1}
\end{equation}
where,
\begin{equation}
{\bf M}=\left(
\begin{array}{cc}
0 & +i\nabla\times \\
-i\nabla\times & 0
\end{array}\right)
\hspace{5mm};\hspace{5mm}
{\bf P}=\left(
\begin{array}{cc}
\varepsilon({\bf r})\varepsilon_{0} & 0 \\
0 & \mu({\bf r})\mu_{0}
\end{array}\right)
\end{equation}
We now define a six vector,
\begin{equation}
{\bf F}_{s}=\left(\begin{array}{c}{\bf E}_{s} \\
{\bf H}_{s}\end{array}\right) 
\end{equation}
as an eigenfunction of equation~\ref{eqn:dos1} with an eigenvalue $\omega_{s}$.
We choose to normalise the ${\bf F}$'s such,
\begin{equation}
\int \sum^{6}_{j=1}\sum^{6}_{j'=1} F_{s,j}^{\dagger}({\bf r}) P_{jj'}
F_{s',j'}({\bf r}) d^{3}{\bf r} = \delta_{ss'}
\end{equation}
And the completeness relation gives us,
\begin{equation}
\sum_{s,j''}F_{s,j}({\bf r})F_{s,j''}({\bf r'})P_{j'j''}=\delta_{jj'}
\delta({\bf r-r'})
\end{equation}
We can define a Green's function in the usual way,
\begin{equation}
G^{R}_{jj'}(\omega,{\bf r},{\bf r'})=\sum_{s,j''}
\frac{F_{j}(s,{\bf r})F^{\dagger}_{j''}(s,{\bf r'})P_{j''j'}}
{\omega-\omega_{s}+i\delta}
\end{equation}
The Green's function clearly obeys the equation,
\begin{equation}
(\omega-{\bf P}^{-1}{\bf M})G^{R}_{ij}(\omega,{\bf r},{\bf r'})
= \delta_{ij}\delta({\bf r-r'})
\end{equation}
We can now Fourier transform to obtain the Green's function in the
time domain,
\begin{eqnarray}
G^{R}_{jj'}(t,{\bf r},{\bf r'})&=&
\frac{1}{2\pi}\int^{+\infty}_{-\infty}G^{R}_{jj'}
(\omega,{\bf r},{\bf r'})\Exp{-i\omega t} \ d\omega \nonumber
\\ &=& -i\sum_{sj''} F_{s,j}({\bf r})
F^{\dagger}_{s,j''}({\bf r'})P_{j''j'}\Exp{-i\omega_{s}t} 
\end{eqnarray}
so that,
\begin{equation}
\left(i\frac{\partial}{\partial t} - {\bf P}^{-1}{\bf M}\right)
{\bf G}^{R}(t,{\bf r},{\bf r'})=\delta(t)\delta({\bf r-r'})
\end{equation}

Now we turn to consider how to repeat this procedure for the discrete
case. We begin from equation~\ref{eqn:dos1},
%\begin{equation}
%\sum_{{\bf k'}}({\bf P}^{-1}{\bf M})_{{\bf kk'}}F_{{\bf k'}}=
%\omega_{{\bf k}}F_{{\bf k}}
%\end{equation}
but apply the substitutions for ${\bf k},\omega$ presented in the
previous section,
\begin{equation}
{\bf P}^{-1}{\bf M}\left(\begin{array}{cc}{\bf E}({\bf r},t)
\\ {\bf H}({\bf r},t)\end{array}\right) =
i\left(\begin{array}{cc} \Delta^{+}_{t} & 0 \\ 0 & \Delta^{-}_{t}
\end{array} \right)\left(\begin{array}{cc}{\bf E}({\bf r},t)\\
{\bf H}({\bf r},t)\end{array}\right) 
\label{eqn:discrete}
\end{equation}
where ${\bf M}$ is now
\begin{equation}
{\bf M}=\left(\begin{array}{cc} 0 & +i\nabla^{-}\times \\
-i\nabla^{-}\times & 0 \end{array}\right)
\end{equation}
%For the eigenmode $s$ this gives,
%\begin{equation}
%{\bf P}^{-1}{\bf M} {\bf F}_{s} = \frac{i}{-i\delta t}
%\left(\begin{array}{cc} (\Exp{-i\delta t \omega_{s}}-1) & 0 \\ 0 & (1 -
%\Exp{i \delta t \omega_{s}}) \end{array} \right)
%{\bf F}_{s}
%\end{equation}
We want to construct the ${\bf G}^{R}(t)$ which will obey the
equation
\begin{equation}
\left[ i\left(\begin{array}{cc} \Delta^{+}_{t} & 0 \\ 0 &
\Delta^{-}_{t} \end{array}\right) - {\bf P}^{-1}{\bf M}\right]
{\bf G}^{R}(t,{\bf r},{\bf r'}) = \delta(t)\delta({\bf r-r'})
\end{equation}
Try constructing,
\begin{eqnarray}
G^{R}_{jj'}(t,{\bf r},{\bf r'}) = -i\delta t \sum_{s,j''}
F_{s,j}({\bf r}) F^{\dagger}_{s,j''}({\bf r'}) P_{j''j'} \Exp{-i
\omega_{s} (t-\delta t)}
&;& t>0 \nonumber \\
G^{R}_{jj'}(t,{\bf r},{\bf r'}) = 0 \hspace{59mm} &;& t \leq 0
\label{eqn:green}
\end{eqnarray}
then,
\begin{eqnarray}
G^{R}_{jj'}(\omega,{\bf r},{\bf r'})  &=&
\sum^{N_{t}\delta 
t}_{t=\delta t} G^{R}_{jj'}(t,{\bf r},{\bf r'}) \Exp{i\omega t}
\\
 &=&
-i\delta t \sum_{s,j''} F_{s,j}({\bf r})
F^{\dagger}_{s,j''}({\bf r'}) P_{j''j'}\left[
\frac{1-\Exp{i(\omega-\omega_{s})N_{t}\delta
t}}{\Exp{-i(\omega-\omega_{s}) \delta t} -1}\right] \Exp{i\omega_{s}
\delta t} \nonumber
\end{eqnarray}
If $N_{t}$ is large and $\omega=\omega+i\delta$,
\begin{equation}
G^{R}_{jj'}(\omega,{\bf r},{\bf r'})=
-i\delta t \sum_{s,j''} F_{s,j}({\bf r})
F^{\dagger}_{s,j''}({\bf r'}) P_{j''j'}
\frac{\Exp{i\omega_{s}\delta t}}
{\Exp{-i(\omega-\omega_{s}) \delta t} -1} 
\end{equation}
which in the limit $\delta t \rightarrow 0$ gives the correct form for
the frequency dependent Green's function. So simply by setting
$t=\delta t$ in equation~\ref{eqn:green} to give,
\begin{equation}
G^{R}_{jj'}(t=\delta t,{\bf r},{\bf r'})=
-i\ \delta t \ \delta_{jj'}\ \delta({\bf r-r'})
\end{equation}
as the appropriate starting condition and applying
equation~(\ref{eqn:discrete}) we can calculate ${\bf G}^{R}(t)$ at
all subsequent times.

Once we have found the Green's function it is a simple matter to
calculate useful physical quantities from it, such as the photonic
density of states. The local density of states, for example, is found
in the usual 
way, from the imaginary part of the trace of $G^{R}(\omega)$~\cite{econ}.
\begin{equation}
\rho(\omega,{\bf r})=-\frac{1}{\pi}{\rm Im}\left[
\sum_{j} G^{R}_{jj}(\omega,{\bf r},{\bf r})
\right]
\end{equation}
Similarly, the band structure can be easily determined by locating
the poles in the Green's function and so identifying the normal modes
of the system.

\section{Results} 

We shall now put the ideas of the previous sections to work for two
different physical systems. 

\subsection{Bragg Stack}
In the first case we will look at what is
probably the simplest one dimensional photonic crystal that can be
imagined, the 
dielectric multilayer, or Bragg stack. This crystal is formed by
stacking together alternating layers of dielectric of high and low
refractive index (see figure~\ref{fig:bragg}). Each layer is of
infinite extent in the plane and we choose parameters such that the
high refractive index ($n=3.6$) planes have a thickness of $0.3a$ and
the low index ($n=1.0$) planes have a thickness of $0.7a$ where $a$ is
the lattice spacing.
This choice leads to a sizable band gaps for electromagnetic waves
propagating normal to the planes as can seen in
figure~\ref{fig:bragg_band}. The frequencies in this figure are scaled
so as to be dimensionless, what we actually write for the frequency is
$\omega a/ (2\pi c_{0})$. The wavevectors are also written in
dimensionless units of $k.a$ so that the edge of the first Brillouin
zone occurs at $\pm\pi$.

In figure~\ref{fig:bragg_dos} we show the local density of
states for a point inside the Bragg stack. Again the frequencies are
given in dimensionless units and the density of states itself is in
arbitrary units. The band gaps are clearly visible as frequency windows
over which the density of states is strongly suppressed. At the band
edges the Van Hove singularities are also apparent corresponding to
the points on the band structure at the band edges where
$\partial\omega/\partial k$ 
tends to zero.

We can go one step further and add a defect layer to the otherwise
perfect crystal. We do this by introducing a super-cell consisting of
25 repeat 
units of the Bragg stack, then a defect of $0.3a$ of the high
refractive index material, and then another 25 unit cells.
The local density of states at the centre of the defect can be seen in
figure~\ref{fig:bragg_defect_dos} superimposed over the density of
states for the perfect crystal. The dominant feature are the new peaks
which have appeared in the band gaps. These correspond to localised
modes associated with the defect. That these modes are tightly
localised can be easily shown by calculating the local density of
states in the crystal several lattice spacings away from the
defect. The perfect crystal result is recovered.

\subsection{Diamond Lattice}

The second system which we shall consider is a fully
three dimensional photonic crystal made up from air spheres embedded
in a dielectric host arranged in a diamond lattice structure. This
structure was one of the first to be shown to have a complete photonic
band gap~\cite{Ho}. 
Figure~\ref{fig:cell} gives a diagram of the diamond structure and
indicates the unit cell which we choose. This unit cell is not
primitive but is chosen for convenience with lattice vectors;
${\bf a}=a{\bf i}$, ${\bf b}=(a/2){\bf i} +
(\sqrt{3}a/2){\bf j}$ and 
${\bf c}=\sqrt{6}a{\bf k}$ where $a$ is the distance between nearest
neighbour lattice points.
The exact parameters we have chosen lead to a
fairly sizable band gap. The refractive index for the host material is
 $n=3.6$, the spheres have $n=1.0$, and the ratio of sphere radius to
the distance separating nearest neighbour spheres is 0.43. The
band structure for this system in the directions $\Gamma - {\rm K}$
and $\Gamma - {\rm X}$ is given in figure~\ref{fig:dia_band}
where it is clear that a band gap is opening up over a
frequency range of about 3.2 and 3.9 in dimensionless units. This band
gap is confirmed in the density of states, in figure~\ref{fig:dia_dos}.
The density of states calculation is performed by creating a
super-cell by stacking unit cells together, eight in each
direction. 
Because each of our unit cells is not in fact primitive but itself
contains three primitive cells, the super-cell consists of 1536
primitive units cells.
By creating a larger unit cell we reduce the number of k-points which
we need to sample in order to build up the complete local density of
states.

Next we can introduce a defect into the diamond lattice by again
creating a super-cell by stacking unit cells together and then cutting
a hole in one of the dielectric parts of the cell.
This defect will
have a localised defect mode associated with it and by altering the
amount of dielectric material we cut away we can tune the frequency of
the mode. We choose the size of the defect so that the defect mode
lies within the band gap for the crystal. 
We cut away a parallelepiped region spanning adjacent unit
cells starting from $({\bf a}/2,{\bf b}/3,{\bf c}/3)$ 
and with edges of $5{\bf a}/6$, $5{\bf b}/6$ and $5{\bf c}/12$.
Figure~\ref{fig:dia_defect_dos} shows the density of states
for the super-cell with the defect and as for the Bragg stack case,
several defect modes are clearly seen. We have also checked that we
we have made the super-cell sufficiently large by testing that the
position of the defect mode peaks do not disperse with ${\bf k}$.

\section{Conclusions}

In this paper we have shown how an Order-N, finite difference time
domain method can be used to calculate the Green's function
in a simple and straightforward way. Specifically, our new formalism
enables us to give a simple derivation for the 
numerical stability criterion, exact statements of charge and energy
conservation and allows us to use non-orthogonal co-ordinate systems
without the usual computational overheads.

From the Green's function a whole
range of other physical quantities can be found such as the local
density of states. The times required to calculate these quantities
scales linearly with the size of the system, so our technique is of
particular importance in analysing systems with very large or
complicated unit cells. Specifically, we have calculated the local
density of states for both a one and three dimensional photonic
crystal containing a defect and have recovered the expected
result. The defect has the effect of introducing a highly localised
mode the frequency of which is determined by the size of the
defect. By a careful choice we have found defect states within the
photonic band gap for the crystal. These localised states have
potential application in photonic cavity laser structures where the
efficiency of the laser is enhanced by suppression of emission into
all but the lasing mode. The linear scaling of this method with system
size should allow modelling of realistic designs for cavity structures
without prohibitive computational overheads associated with
traditional computational schemes.

The computer program used to calculate the results presented in this
paper has recently been submitted to the CPC International Program
Library~\cite{onyx}.
\bibliographystyle{prsty}
\bibliography{DOS}
\begin{figure}[tb]
% \resizebox{!}{7cm}{\includegraphics{bragg.eps}}
 \caption{A dielectric multilayer or Bragg stack formed by alternating
layers of high (shaded) and low (unshaded) permittivity materials. The
thickness of the high dielectric layer is $d$ and the lattice period
is $a$}
 \label{fig:bragg}
\end{figure}
\begin{figure}[tb]
% \resizebox{!}{10cm}{\includegraphics{bragg_band.eps}}
 \caption{The photonic band structure for Bragg stack}
 \label{fig:bragg_band}
\end{figure}
\begin{figure}[tb]
% \resizebox{!}{10cm}{\includegraphics{bragg_dos.eps}}
 \caption{The density of states for the ideal Bragg stack. The
Van Hove singularities in the perfect crystal are clearly seen}
 \label{fig:bragg_dos}
\end{figure}
\begin{figure}[tb]
% \resizebox{!}{10cm}{\includegraphics{bragg_defect_dos.eps}}
 \caption{The density of states both with and without a defect
inserted into the ideal Bragg stack. Both the Van Hove singularities in
the perfect crystal and the localised state associated with the defect
are clearly seen}
 \label{fig:bragg_defect_dos}
\end{figure}
\begin{figure}[tb]
% \resizebox{!}{10cm}{\includegraphics{tri_latt.eps}}
 \caption{Diagram of the diamond lattice. The bravais lattice is FCC and
the lattice points are shown as the black dots. On each lattice point
are placed two `atoms' - dielectric spheres in our case; one at
 $(0,0,0)$, one at $(0,0,c/4)$, where $c=\protect\sqrt{6} a$. Some of
the `atoms' 
are shown as the grey circles. The unit cell we model is the
parallelepiped enclosed by the thick black lines}
 \label{fig:cell}
\end{figure}
\begin{figure}[tb]
% \resizebox{!}{10cm}{\includegraphics{dia_band.eps}}
 \caption{Partial band structure for spheres on the diamond
lattice in the \protect$\Gamma - {\rm K}$ and \protect$\Gamma -
{\rm X}$ directions. The presence of the band gap is clearly shown}
 \label{fig:dia_band}
\end{figure}
\begin{figure}[tb]
% \resizebox{!}{10cm}{\includegraphics{dia_dos.eps}}
 \caption{The density of states for the diamond structure}
 \label{fig:dia_dos}
\end{figure}
\begin{figure}[tb]
% \resizebox{!}{10cm}{\includegraphics{dia_def_dos.eps}}
 \caption{The density of states both with and without a defect
for the diamond structure. The localised state associated with the defect
is clearly seen}
 \label{fig:dia_defect_dos}
\end{figure}
\end{document}